\documentclass[aoas,preprint]{imsart2}
\pdfoutput=1

\RequirePackage[OT1]{fontenc}
\RequirePackage{amsthm,amsmath}
\RequirePackage{natbib} 
\usepackage{graphicx}    
\usepackage{amssymb} 
\usepackage{subfigure}

	\newcommand{\bfmu}{\mbox{\boldmath $\mu$}}
	\newcommand{\bfPsi}{\mbox{\boldmath $\Psi$}}
	\newcommand{\bfSigma}{\mbox{\boldmath $\Sigma$}}

\begin{document}

% ---- TITLE ---------->
\begin{frontmatter}
\title{Spatio-temporal Modelling of Temperature Fields in the Pacific Northwest}
\runtitle{Temperature Fields in the Pacific Northwest} 

\begin{aug}
\author{\fnms{Camila} \snm{Casquilho-Resende}\thanksref{m1, t1}\ead[label=e1]{camila.casquilho@stat.ubc.ca}},
\author{\fnms{Nhu D.} \snm{Le}\thanksref{m1,m2}\ead[label=e2]{nle@bccrc.ca}}
\and
\author{\fnms{James V.} \snm{Zidek}\thanksref{t2,m1}\ead[label=e3]{jim@stat.ubc.ca}
\ead[label=u1,url]{http://www.stat.ubc.ca}}

\thankstext{t1}{Corresponding author}
\thankstext{t2}{The work reported in this manuscript was supported by a grant from the Natural Sciences and Engineering Research Council of Canada.}
\runauthor{Casquilho-Resende, Le and Zidek}

\affiliation{Department of Statistics, The University of British Columbia \thanksmark{m1} and BC Cancer Agency, British Columbia, Canada \thanksmark{m2}}

\address{C. M. Casquilho-Resende \\
J. V. Zidek \\
Department of Statistics \\
The University of British Columbia \\
3182 Earth Sciences Building \\
2207 Main Mall \\
Vancouver, BC, V6T 1Z4 \\
Canada \\
\printead{e1}\\
\phantom{E-mail:\ }\printead*{e3}}
\address{N. D. Le \\
BC Cancer Agency\\
675 West 10th Avenue \\
Vancouver, BC, V5Z 1L3 \\
Canada
}
\end{aug}

% ---- Abstract ---------->

\begin{abstract}
The importance of modelling temperature fields goes beyond the need to understand a region's climate and serves too as a starting point for understanding their socioeconomic, and health consequences. The topography of the study region contributes much to the complexity of modelling these fields and demands flexible spatio-temporal models that are able to handle nonstationarity and changes in trend. In this paper, we develop a flexible stochastic spatio-temporal model for daily temperatures in the Pacific Northwest, and describe a methodology for performing Bayesian spatial prediction. A novel aspect of this model, an extension of the spatio-temporal model proposed in \cite{Le1992}, is its incorporation of site-specific features of a spatio-temporal field in its spatio-temporal mean. Due to the often surprising Pacific Northwestern weather, the analysis reported in the paper shows the need to incorporate spatio-temporal interactions in that mean in order to understand the rapid changes in temperature observed in nearby locations and to get approximately stationary residuals for higher level analysis. No structure is assumed for the spatial covariance matrix of these residuals, thus allowing the model to capture any nonstationary spatial structures remaining in those residuals. 
\end{abstract}

% ---- End Abstract ------->


\end{frontmatter}

% ---- End Front Matter ---------->

% ---- PAPER BEGINS  ---------->

\section{Introduction}

Meteorological variables are crucial to understand a region's climate. In particular, a much discussed topic in recent years is that the earth's climate has been changing: global average and sea surface temperature have increased and extreme temperature events such as heat waves are now more frequent. This changing climate has led to concerns about its impact on human health. 

Extreme temperatures may contribute to cardiovascular and respiratory diseases, especially among elderly people, as it can be seen in \cite{Astroem2013}. \cite{Li2012} studied the relationship between temperature and morbidity due to extreme heat and revealed that a number of hospital admissions in Milwaukee, Wisconsin were detected to be significantly related to high temperature. In fact, \cite{Robine2008} estimates an excess death toll of 70,000 people due to high temperatures in Europe in 2003 and a World Health Organization (WHO) assessment concluded that the modest warming that has occurred since the 1970s was already causing over 140,000 excess deaths annually by the year 2004 \citep{WHO2009}. The spread of infectious diseases is also now being linked to the climate change as per \cite{Hoberg2015}. All of this highlights that the importance of modelling temperature fields goes well beyond the natural sciences. 

This paper focuses on the Pacific Northwest and the period January to June of the year 2000. That is because its genesis lies in earlier work for that region and time period \citep{Liu2007} aimed at developing a extension of Bayesian melding for ensembles of metereological models for temperature forecasting, work that could not be completed due the extreme nonstationarity of temperature fields in that region. A completion of that work will be presented in a sequel to this paper, which lays the required foundation for modelling the temperature field. However, the theory and modelling approach are general and we expect them to be applicable to other regions and time periods. %JZ

The Pacific Northwest is the region in the western part of North America adjacent by the Pacific Ocean. In this paper, we develop a stochastic spatio-temporal model for daily temperature in this region. We also develop a methodology for performing spatial prediction, referred to as Bayesian spatial prediction. Our methodology is an extension of the spatio-temporal model proposed in \cite{Le1992} by allowing the inclusion of site-specific features in the spatio-temporal mean. This is extremely important due to the often rapid changes in temperature trends in the Pacific Northwest, surprisingly even for nearby locations. 

The Pacific Northwest is a rather diverse region, with four mountain ranges dominating it, including the Cascade Range, the Olympic Mountains, the Coast Mountains and parts of the Rocky Mountains. This region is known to have a wet, cool climate overall, though in more inland areas, climate can be fairly dry, with warmer summers and harsher winters.  According to \cite{Mass2008}, the Northwest weather and climate are dominated mainly by the Pacific Ocean to the west and the region's mountain ranges that block and deflect low-level air. \cite{Mass2008} notes that the ocean moderates the air temperatures year-round and serves as a source of moisture, and the mountains modify precipitation patterns and prevent the entrance of wintertime cold-air from the continental interior. 

The terrain is another key element to understand the Pacific Northwest weather. East of the Rocky Mountains is where usually the coldest air locates itself, but the Rockies protect this cold air from reaching the Northwest and the part that manages to do so gets warmer when descending to eastern Washington, Oregon and the Cascade Range. The temperatures in this region are thus mainly controlled by the proximity to the Pacific Ocean and by elevation.

In the literature, spatial modelling in the Pacific Northwest is also recognized to be rather complex and it has been the subject of critical observation by local weather scientists. More recently, \cite{Mass2008} apprises that the weather in the Northwest is often surprising, both in its intensity and in the remarkable contrasts between nearby locations. Rapid changes and localized weather are very common in this region and the terrain plays an important role in separating often radically different climate and weather regimes. \cite{Mote2004} performed an assessment of the impact of the global climate change in the Pacific Northwest, by using station data collected between 1920 and 1997 and noticed an apparent tendency for high-elevation stations to exhibit weaker warming trends than lower-elevation stations when examining temperature trends in this region. 

Especially due to the topography of the study region, the modelling of temperature fields can be quite challenging. \cite{Kleiber2013} recognized the difficulty faced by statistical models in capturing the complex spatial variability. By analyzing data from the state of Colorado, \cite{Kleiber2013} developed a bivariate stochastic temperature for minimum and maximum temperature via a nonparametric approach. In the Pacific Northwest, \cite{Salathe2008} focused on the development of a regional climate model run at a 15-km grid spacing. 

One of our contributions is the ability to accommodate features in the mean that vary on space by extending the spatio-temporal model proposed in \cite{Le1992}, and easily performing spatial prediction. The methodology is described in Section \ref{sec:BSP}, and another important feature is its flexibility due to the fact that no structure is assumed for the spatial covariance matrix. The method thus is able to accommodate nonstationarity. We illustrate our analysis based on the \cite{Sampson1992} method for estimating nonstationary spatial covariance structure. 

In this paper, we deal with station data collected from U.S. Global Historical Climatology Network. Due to the remarkable existence of a high spatial correlation at widely separated sites, we address the need for interactions in the spatio-temporal mean.  We also consider averaged temperature values over a 30-year (1981-2010) range, used as a proxy for the mean field. This also enables us to directly model what could not have been explained by the expected climate. These data was obtained using a state of the art climate model called PRISM (Parameter-elevation Relationship on Independent Slopes Model), described in \cite{Daly1994, Daly1997, Daly2000}. We provide an overview of the PRISM in Section \ref{PRISM}. Furthermore, the prediction performance of our method is also compared with the ordinary kriging of the residuals, after taking into account the spatio-temporal mean effects. 

In the following Section \ref{DataDescription}, we provide a description of the data and some interesting features to understand the weather in the Pacific Northwest. The methodology is described in Section \ref{sec:BSP}, and finally, in Section \ref{sec:Results}, we describe our results and perform a model assessment. 

\section{Data Description and Preliminary Analyses}\label{DataDescription}

\subsection{U.S. Global Historical Climatology Network}\label{NCDC}

The U.S. Global Historical Climatology Network - Daily (GHCND) is an integrated database of climate summaries from land surface stations across the globe, developed for several potential applications, including climate analysis and studies that require data at a daily time resolution, as described in \cite{Menne2012} and \cite{Lawrimore2011}. 

\begin{figure}[h!]
	\centering
	\begin{tabular}{cc}
		\includegraphics[scale=0.35]{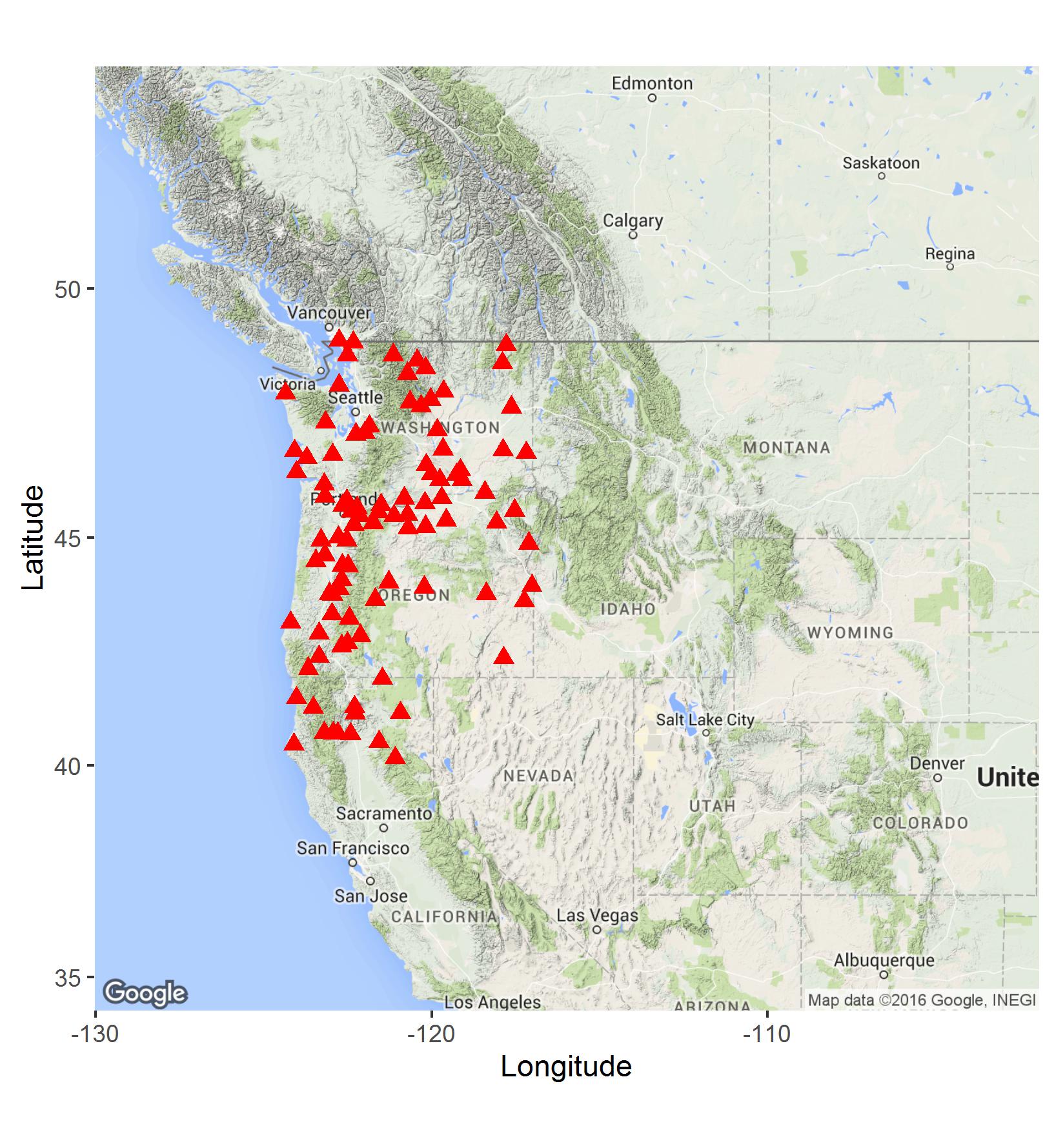}
	\end{tabular}
	\caption{Locations of 97 stations in the Pacific Northwestern area considered in this study.}
	\label{fig:AllStationsMap}
\end{figure}

Figure \ref{fig:AllStationsMap} illustrates the locations of 97 stations where maximum daily temperature data were downloaded from the GHCND database for the investigation reported in this paper. The maps images displayed in this paper were obtained via the \texttt{ggmap R} package \citep{RCT2014, Kahle2013}.

For the reasons described in the introduction, the selected spatio-temporal data were for the January to June 2000 period. We thus emulate the 48-hour forecasts of surface level temperature data (initialized at midnight Coordinated Universal Time) available at the University of Washington (UW) Probcast Group web page (\url{http://www.stat.washington.edu/MURI/}). 

That were to be the subject of our initial investigation until we learned that it is not an integrated spatio-temporal data set in as much as the locations at which measurements are available differ considerably from one day to the next--some stations have very few measurements and the temporal spacing of the observations is highly irregular.

\begin{figure}[h!]
	\centering
	\begin{tabular}{cc}
		\includegraphics[scale=0.53]{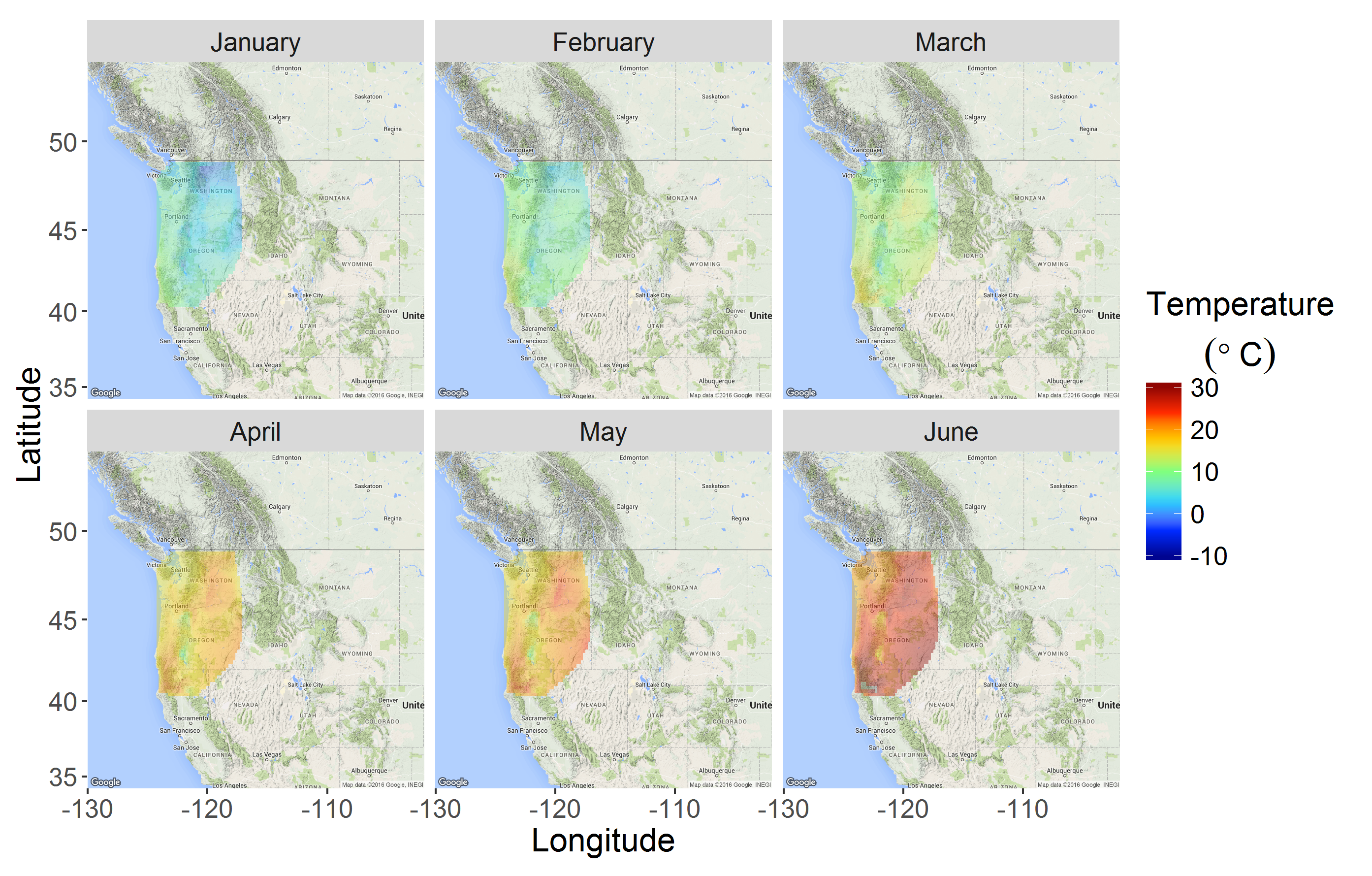}
	\end{tabular}
	\caption{Averaged site temperatures for different months. Notice the different patterns of temperature variation across the region.} \label{mapMonthlyImagesNCDC}
\end{figure}

\newpage
Figure \ref{mapMonthlyImagesNCDC} shows contours of average site temperatures for different months, obtained by bivariate linear interpolation. Notice that cooler temperatures are observed closer to the Pacific Ocean. Another interesting feature is the different patterns of temperature variation across the region. Warmer temperatures are generally found east of the Cascades and since western Washington is more exposed to air coming from from Puget Sound, the Straits of Juan de Fuca and Georgia, and the Pacific Ocean, it generally experiences cooler temperatures.

We start our preliminary analysis an exploration of the spatio-temporal trend, followed by an analysis of the unexplained residuals of the spatio-temporal process. Initially a simple geostatistical model was considered where the spatial trend is described through a second-order polynomial regression model. The temperature measured on day $t$ at location $s$ is denoted as $Y_t(s)$, where $t$ denotes the time in days and ${\bf s} = (s_1; s_2)$,  the location coordinates, in km, after a suitable projection of the relevant part of the globe onto a flat surface. We considered projected spatial coordinates using the Lambert conformal conic projection, but for simplicity, we still refer to these projected coordinates as simply latitude and longitude. For a fixed time $t$, 
\begin{eqnarray}
Y_t({\bf s}) &=& \mu({\bf s}) + \nu({\bf s}), ~ \nu({\bf s}) \sim N(0,\sigma^2) \\
\mu({\bf s}) &=& \alpha + \beta_{1} s_1 + \beta_{2} s_2 + \beta_{3} s_1s_2 + \beta_{4} s_1^2 +  \beta_{5} s_2^2. \nonumber
\end{eqnarray}

Recall that when the spatial random field is stationary, the semivariogram between two locations ${\bf s}_{k}$ and ${\bf s}_{l}$ at a fixed time point $t$ is defined as
\begin{eqnarray}
\gamma({\bf s}_{k}, {\bf s}_{l}) = \frac{1}{2} E[(Y_t({\bf s}_k)-Y_t({\bf s}_l))^2].
\end{eqnarray} 

\begin{figure}[h!]\begin{center}\vspace{-0.3cm}  
\begin{tabular}{cc}
\includegraphics[scale=0.5]{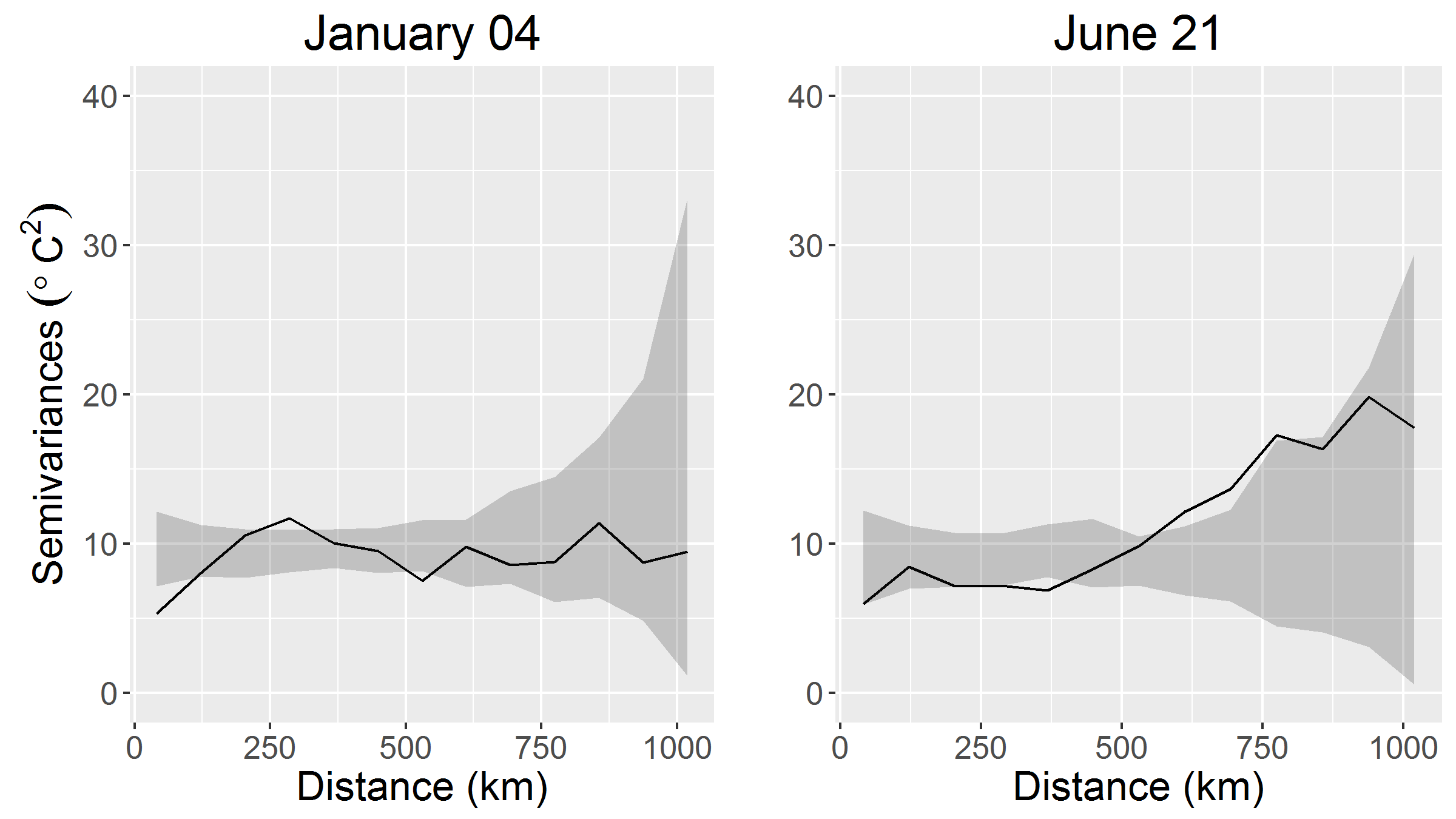}
\end{tabular}    
\caption{Binned empirical semivariograms with Monte Carlo envelopes in shaded area. These envelopes are obtained by permutations of the data across the sampling locations, and indicate regions of uncorrelated data. }\label{variogBinNCDC}
\vspace{-.69cm}
\end{center} 
\end{figure}

For illustrative purposes, Figure \ref{variogBinNCDC} contains the binned empirical semivariogram obtained separately for each of the two selected days (January 04 and June 21). The shaded area corresponds to Monte Carlo envelopes obtained by repeatedly recomputing and plotting the semivariance after permutations of the temperature data across the sampling locations. They indicate regions of uncorrelated data. 

\begin{figure}[h!]
	\centering
	\includegraphics[scale = 0.5]{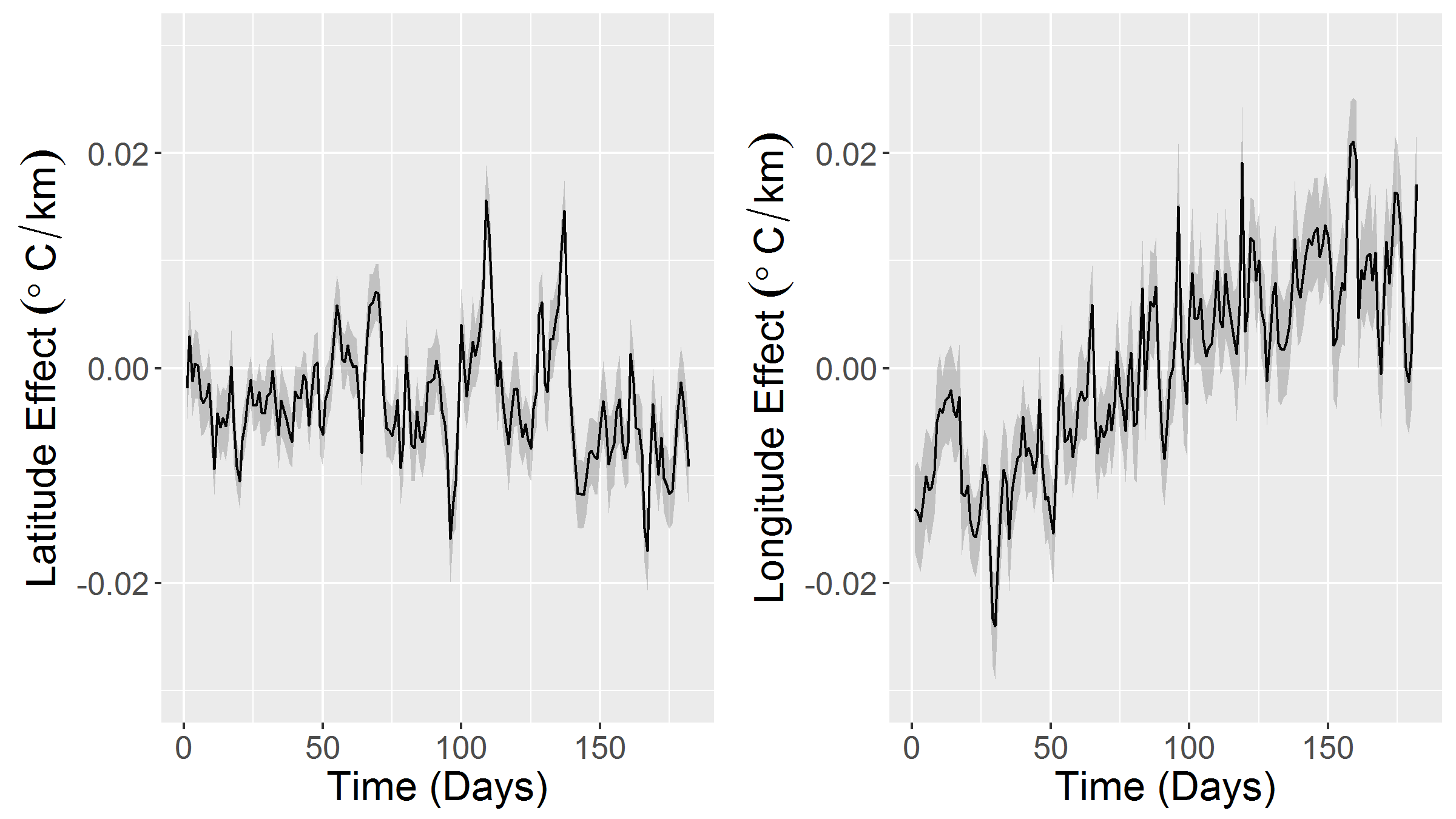}
	\caption{Latitude and longitude effects changing over time. The shaded area represents 95\% confidence intervals for these effects.}
	\label{latlongeffNCDC}
\end{figure}

Figure \ref{latlongeffNCDC} illustrates how the latitude and longitude effects change over time. The longitude effect has a clearly increasing trend and this is possibly due to the Cascade mountains that extend from southern British Columbia through Washington and Oregon to Northern California. The Cascades block the westward movement of most of the cold, dense air that manages to reach eastern Washington and Oregon. 

Our preliminary analysis indicates that for regions where topography changes significantly, simple polynomial trends commonly used may introduce bias in the spatio--temporal residuals resulting in a semivariogram with a large squared bias term that can lead to a spurious finding of nonstationarity when none exists. Thus, our preliminary analysis points to the need for the improved estimation of the spatial mean model as reported in the sequel, one that accounts for extra features, notably elevation. 

\newpage In particular, we recognize that our analysis needs to include spatio--temporal interactions as well as longitude--elevation interaction. The latter is due to the effect of the proximity to the Pacific Ocean, which also takes into consideration the elevation effect due to the mountain ranges when moving eastward. Moreover, the longitude effect is assumed to depend on the elevation as well as how far north the station is located. And this effect must be allowed to vary over time. Similarly, the latitude effect may vary over time and it is dependent on how far east the weather station is located. 

The above considerations lead to a spatio--temporal mean (trend) function that may be described as follows:
\begin{eqnarray}
\mu_t({\bf s}) &\equiv& f(\texttt{\mbox{long}*\mbox{lat}*\mbox{month},\mbox{long}*\mbox{elev}}), \label{MeanFct}
\end{eqnarray} where $f$ denotes a linear function, ${\bf s} = (\texttt{long}, \texttt{lat})$ are projected latitude and longitude coordinates, $t$ indicates the month in study, and \texttt{elev} the elevation at ${\bf s}$. The $*$ notation is used to indicate that the mean function includes the individual, the two-way and, where applicable, the three-way interaction effects. 

However an alternative suggests itself, one based on the use of historical temperature averages over the region to account for these complex interactions, as a representation of the climate in the Pacific Northwest. We describe this alternative in the following Subsection \ref{PRISM}.

\subsection{PRISM Climate Group Data} \label{PRISM} %-------------->

PRISM is a climate analysis system that uses point data, a digital elevation model (i.e. digital representations of cartographic information in a raster form), and other spatial data to generate gridded estimates of annual, monthly and event-based climatic parameters \citep{Daly1994, Daly1997}. It was developed primarily to interpolate climate elements in physiographically complex landscapes \citep{Daly2008}, and is particularly useful to identify short and long-term climate patterns.

The extrapolation of climate over high elevation ranges is often needed due to the lack of observations in mountainous regions. The use of PRISM data would then be ideal for complex regions with mountainous terrain such as the Pacific Northwest. In the literature, \cite{Daly1994, Daly1997, Daly2000} provide a description of the methodology behind PRISM. The main idea is that it calculates linear parameter--elevation relationships, allowing the slope to change locally with elevation. Observations nearer to the target elevation receive more weight than those further up or down slope. 

For temperature, the elevation of the top of the boundary layer is estimated by using the elevation of the lowest DEM pixels in the vicinity and adding a climatological inversion height to this elevation. \citep{Daly1997} 

The PRISM data we obtained corresponds to average values for temperature computed over a 30-year range (1981-2010), provided by the PRISM Climate Group, Oregon State University, and available online at \url{http://prism.oregonstate.edu}. Our goal is to use these data as a representation of the climate in the Pacific Northwest. Having this information enables a comparison with our observations and an analysis of anomalies (i.e. differences between actual and expected values via PRISM), which would highlight what could not have been explained by the expected climate. This will also serve as a baseline comparison between the more complex mean function proposed in Subsection \ref{DataDescription} that includes spatio--temporal interactions. Finally, in the sequel, the PRISM data is used construct a spatio--temporal trend model as an alternative to that suggested by the analysis reported in Subsection \ref{DataDescription}. 

\section{Bayesian Spatial Prediction}\label{sec:BSP}

In this section, we present an empirical Bayesian spatial prediction (BSP) method built on the assumption that realizations of an underlying random field are obtained from measurements made at $g$ gauged stations and that the goal is to obtain spatial predictions at the other $u$ ungauged stations. Let ${\bf Y}_t \equiv ({\bf Y}_t^{(u)}, {\bf Y}_t^{(g)})$ denote a $p$-dimensional row vector $(p = u + g)$, where ${\bf Y}_t^{(u)}$ and ${\bf Y}_t^{(g)}$ corresponds to the row vectors at the ungauged and gauged stations, respectively. The variables ${\bf Y}_t$ are assumed to be independent over time, or have passed a pre-filtering preliminary step, such that for $t = 1, \dots, n$, 
\begin{eqnarray}\label{eq:bsp}
{\bf Y}_t |{\bf z}_t, {\bf B}, \bfSigma &\sim& \mathcal{N}_{p}({\bf B} {\bf z}_t , \bfSigma), 
\end{eqnarray} where $\mathcal{N}$ denotes a multivariate normal distribution, with subscripts making the dimension explicit; the ${\bf z}_t$ is a $k$-dimensional column vector of covariates and ${\bf B}$ denotes a $(p \times k)$ matrix of regression coefficients. 

As originally formulated \citep{Le1992} in the BSP, covariates were allowed to vary with time, but not space. Over the ensuing decade the BSP was extended in a variety of ways as summarized in \cite{Le2006}. In particular, the response vector at each space--time point could be multivariate, thus enabling site specific random covariates with a Gaussian distribution to be incorporated in the BSP by first including them in the fitted multivariate joint distribution in Equation (\ref{eq:bsp}) and then conditioning on them to get the BSP. However, no way was found to incorporate site specific nonrandom covariates.

However, such covariates are confronted in our analysis of temperature fields in complex regions, as the spatio-temporal mean function must include, say, topographic features as well as the oftentimes crucial spatio-temporal interactions. Thus, an extension was needed and the one that was developed, will now be presented.  
%JZ

Let ${\bf Y}$ be a $(n \times p)$ response matrix such that ${\bf Y} \equiv ({\bf Y}_1, \dots, {\bf Y}_n)$, ${\bf Z}$ is a $(n \times k)$ design matrix and ${\bf B}$ a $(k \times p)$ matrix of regression coefficients. Assume that 
\begin{eqnarray}
{\bf Y} |{\bf Z}, {\bf B}, \bfSigma &\sim& \mathcal{MN}_{n \times p}({\bf Z}{\bf B} , {\bf I},  \bfSigma) \\
{\bf B}| {\bf B}_0, \bfSigma, {\bf F} &\sim&  \mathcal{MN}_{k \times p}({\bf B}_0, {\bf F}^{-1}, \bfSigma) \\
\bfSigma &\sim& \mathcal{W}^{-1}_{p}(\bfPsi, \delta),
\end{eqnarray} where ${\bf F}^{-1}$ is a positive $(k \times k)$ definite matrix, and $\bfPsi$ a $(p \times p)$ hyperparameter matrix. Here, $\mathcal{MN}$ and $\mathcal{W}^{-1}$ denote the matrix normal and the inverted Wishart distributions, respectively, with subscripts making the dimensions explicit. We write

\begin{eqnarray}
{\bf B}_0 = \left( \begin{array}{cccc} \beta_0^{(1)} & \dots & \beta_0^{(p)} \\
\beta_1^{(1)} & \dots & \beta_1^{(p)} \\
\vdots & & \vdots \\
\beta_{k-1}^{(1)} & \dots & \beta_{k-1}^{(p)}
\end{array} \right),
\end{eqnarray} where $\beta_0^j = \alpha + \sum_{l} \beta_{z_l} z_{l_j}$ includes the site-specific covariates at site $j$, denoted as $z_{l_j}$, $j = 1, \dots, p$ and for $i = 1, \dots, k$, $\beta_i^j$ denotes the coefficients of the non-site specific covariates. The first column of the $Z$ matrix corresponds to a unit column vector, whereas the subsequent columns would contain the non--site specific covariates.

Denoting the matrices $\bfSigma_{gg}$ and $\bfSigma_{uu}$ as the covariance matrices of  ${\bf Y}^{(g)}$ and ${\bf Y}^{(u)}$, respectively, and $\bfSigma_{ug}$ the cross-covariance, we can partition $\bfSigma$ and similarly the hyperparameter matrix $\bfPsi$ as  
\begin{eqnarray}
\bfSigma = \left( \begin{array}{cc} \bfSigma_{uu} & \bfSigma_{ug} \\ \bfSigma_{ug} & \bfSigma_{gg} \end{array} \right) ~ \mbox{and} ~ \bfPsi = \left( \begin{array}{cc} \bfPsi_{uu} & \bfPsi_{ug} \\ \bfPsi_{ug} & \bfPsi_{gg} \end{array} \right).
\end{eqnarray}

For a fully (proper) Bayesian approach, extra hierarchy levels could be specified. Nonetheless, the BSP was developed from its inception to save computational time by bypassing this approach. It was recognized that, in practice, the lack of prior knowledge would inevitably lead to a somewhat arbitrary choice of a convenience prior in this high-dimensional model. Thus, a preliminary empirical Bayes step is required for estimating $B_0$ via a linear regression modelling approach, as suggested by the preliminary analysis in Section \ref{DataDescription}.

When performing spatial prediction, we use the result showed in \cite{Le1992} that the conditional distribution of ${\bf Y}_t^{(u)}$, where $t \in \{1, \dots, n\}$ is given by
\begin{eqnarray}\label{eq:bsp}
{\bf Y}_t^{(u)} | {\bf y}_t^{(g)}, {\bf Z}, {\bf B}_0  \sim t_{u} \left(\bfmu^{(u)}, \frac{d}{\delta - u + 1} \bfPsi_{u|g}, \delta - u + 1\right),
\end{eqnarray} where
\begin{eqnarray}
\bfmu^{(u)} &=& {\bf z}_{t}{\bf B}_0^{(u)} + \bfPsi_{ug}\bfPsi_{gg}^{-1}({\bf y}_t^{(g)} - {\bf z}_{t} {\bf B}_0^{(g)}) \\
d &=& 1 + {\bf Z} {\bf F}^{-1} {\bf Z}^{\top} + ({\bf y}_t^{(g)} - {\bf z}_t {\bf B}_0^{(u)})\bfPsi_{gg}^{-1}({\bf y}_t^{(g)} - {\bf z}_t {\bf B}_0^{(u)})^{\top} \\
\bfPsi_{u|g} &=& \bfPsi_{uu} - \bfPsi_{ug}\bfPsi_{gg}^{-1} \bfPsi_{gu}.
\end{eqnarray} Here ${\bf B}_0$ was partitioned as ${\bf B} = ({\bf B}_0^{(u)}, {\bf B}_0^{(g)})$ according to the partition of ${\bf Y}_t$ (superscripts denoting the ungauged and gauged parts).

To finish model development, the covariance of the residual responses in Equation (\ref{eq:bsp}),  $\bfPsi_{u\mid g}$ , must be specified. However, in practice and in our applications, these residuals will not have a second order stationary distribution. Thus we need to handle residuals with a non-stationary distribution and a method for doing so is described in the following section.

\subsection{Handling Nonstationarity}\label{sec:HandleNonstat}

In environmental applications, it is crucial that a spatio-temporal is able to handle nonstationarity. In our work we adopt the celebrated Sampson-Guttorp (SG)  warping method \citep{Sampson1992}, which for completeness, we now describe.  The key idea in the SG approach is that 
of deforming the geographical or ${\cal G}$-space into another, dispersion ${\cal D}$-space, on which domain the process may be considered approximately stationary. The spatial deformation approach then models the spatial covariance as 
\begin{eqnarray}
Cov(Y({\bf s}_i), Y({\bf s}_j)) = 2\rho_{\theta}(||f({\bf s}_i) - f({\bf s}_j)||),
\end{eqnarray} for two locations ${\bf s}_i$ and ${\bf s}_j$ in the study region, where $f$ is a smooth nonlinear map $\mathcal{G} \rightarrow \mathcal{D}$ from the geographical $\mathcal{G}$-space $(\mathcal{G} \subset \mathbb{R}^d)$ to the deformed $\mathcal{D}$-space $(\mathcal{G} \subset \mathbb{R}^{d})$. For notation simplicity, we are omitting the time subscript.

The locations of the sites in the $\mathcal{D}$-space are obtained via a multidimensional scaling algorithm. A mapping of the sites from the $\mathcal{G}$-space into the $\mathcal{D}$-space is obtained by solving the  minimization problem with following criterion, over all monotonic functions $\delta$:
\begin{eqnarray}
\min_{\delta} \frac{\sum_{i<j} [\delta(d_{ij})-h_{ij}]^2}{\sum_{i<j} h_{ij}^2},
\end{eqnarray} where $d_{ij}$ and $h_{ij}$ denote the observed dispersion and the distance between between sites $i$ and $j$ in the $\mathcal{D}$-space, respectively. 

Once the locations of the sites are obtained in the $\mathcal{D}$-space, \cite{Sampson1992} use thin plate splines to obtain a smooth mapping of the sites from the $\mathcal{G}$-space into the $\mathcal{D}$-space and the $\delta$ function is replaced by a smooth function $g$ such that $d_{ij} \approx g(h_{ij})$. It is then possible to obtain estimates of realizations of the spatial process at ungauged locations by first smoothly mapping them onto the $\mathcal{D}$-space and subsequently using standard stationary modeling tools. The smoothness is enforced through a smoothing spline dependent on a parameter $\lambda$ to avoid the overfitting of the dispersions when the analysis results in a folded map.  This is the approach taken in the sequel where we apply our extended BSP method.

\section{Results}\label{sec:Results}

\begin{figure}[h!]
	\centering 
	\begin{tabular}{cc}
		\includegraphics[scale=0.55]{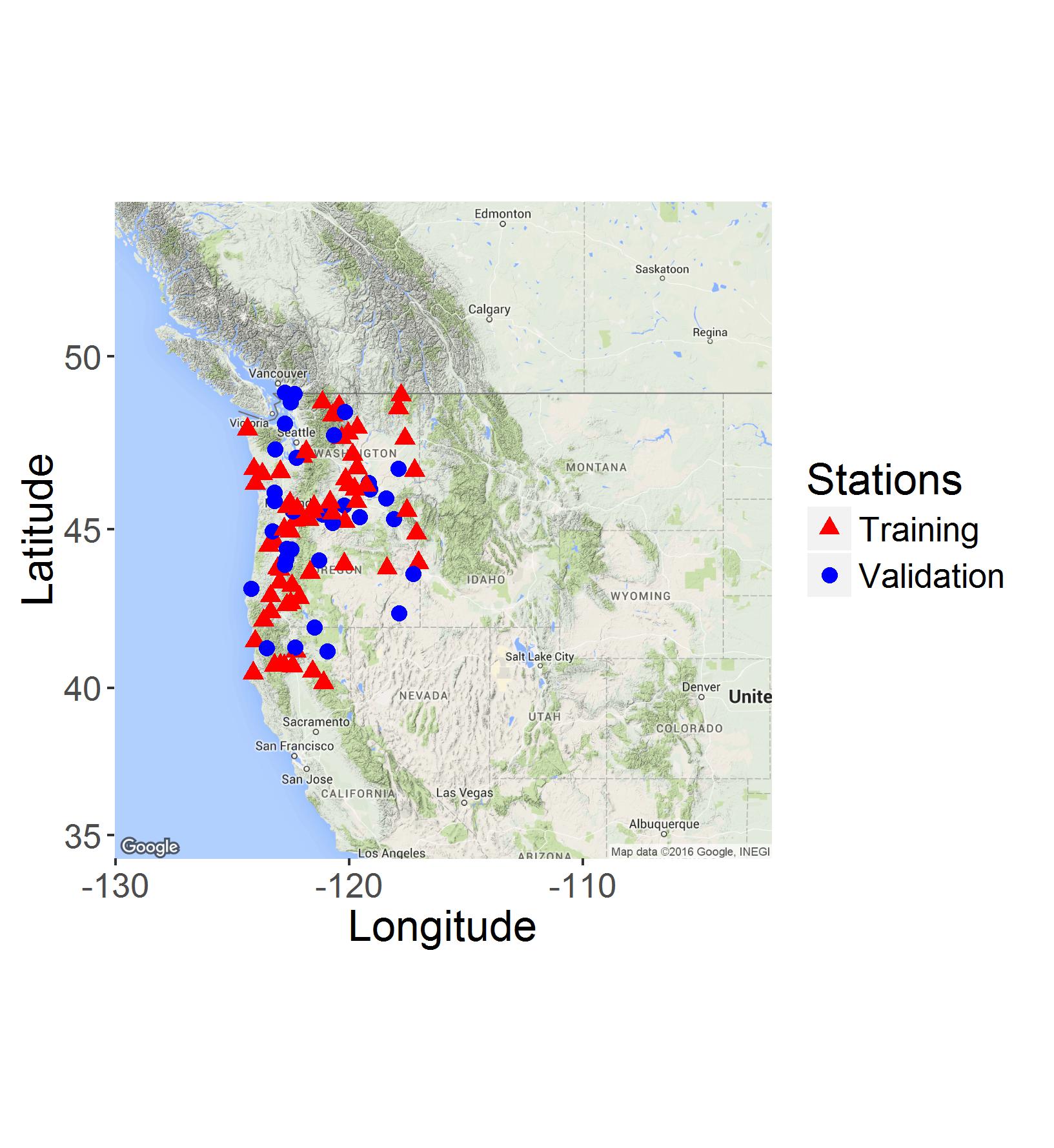}
	\end{tabular} \vspace{-1cm}
	\caption{Locations of the stations selected for training and for validation purposes. }
	\label{fig:AllStationsMapTrainValid}
\end{figure}

This section presents the results of applying the BSP method described in Section \ref{sec:BSP}, implementable using the \texttt{EnviroStat v0.4-0 R} package \citep{Le2014a}. For that purpose, we initially selected 64 stations at random for training, leaving the remainder of the 97 stations for validation purposes, as illustrated in Figure \ref{fig:AllStationsMapTrainValid}.

Work begins with an analysis of the spatio-temporal trend, as it's described in the following Subsection \ref{sec:spt-trend}, followed by an analysis of the spatial correlation in the residuals, after taking into account this trend in Subsection \ref{sec:spatcore}.

\subsection{The spatio--temporal trend} \label{sec:spt-trend}

\begin{figure}[h!]
	\centering
	\begin{tabular}{c}
		\includegraphics[scale=0.35]{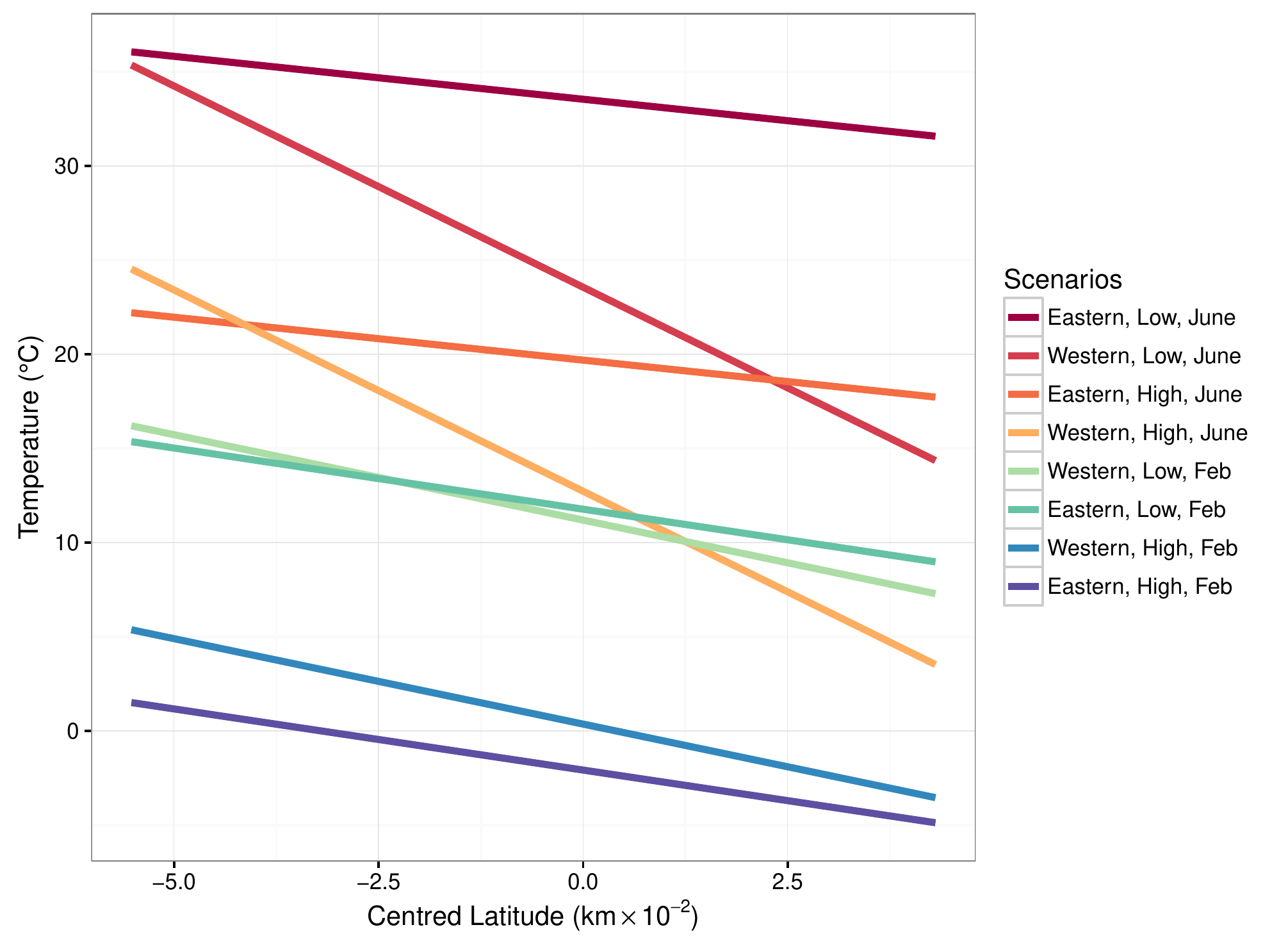}
	\end{tabular}
	\caption{Investigating effect of projected latitude (centred) on temperatures considering different interaction scenarios for longitude (eastern, western), elevation (high, low) and time (February, June).}
	\label{fig:InteractionPlot}
\end{figure}

For the training stations, Figure \ref{fig:InteractionPlot} illustrates the effect of projected latitude on temperatures considering different scenarios of longitude, elevation and time. Notice that moving north implies that the temperature in fact decrease in different rates, depending on your initial scenario. We refer to this as the RC-effect, which refers to a phenomenon where the effect of latitude and longitude changes over time, that is, where at certain times, similar observations may obtain at widely separated sites. In a statistical model, this effect alerts to the need to include space-time interactions. 

\begin{figure}[h!]
	\centering
	\begin{tabular}{c}
		\includegraphics[scale=0.4]{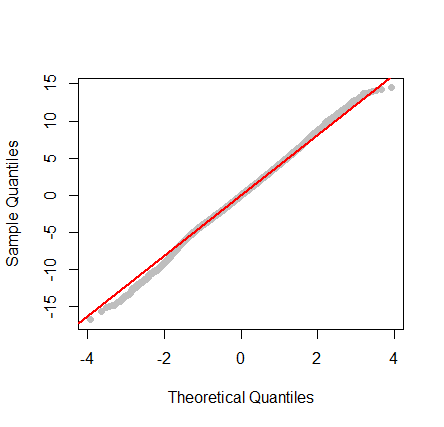}
	\end{tabular}
	\caption{Normal quantile-quantile plot of the residual temperatures of a linear model with spatio-temporal mean function as in Equation \ref{MeanFct}. }
	\label{fig:QQPlotRes}
\end{figure}

For validating the Gaussian assumption, Figure \ref{fig:QQPlotRes} indicates that it may be reasonable without any transformation.

The following Subsection \ref{sec:spatcore} provides an analysis of the spatial correlation in the residuals, after taking into account this spatio-temporal trend.

%\newpage
\subsection{Spatial correlation in the residuals}\label{sec:spatcore}

\begin{figure}[h!]
	\centering
	\begin{tabular}{c}
		\includegraphics[scale=0.4]{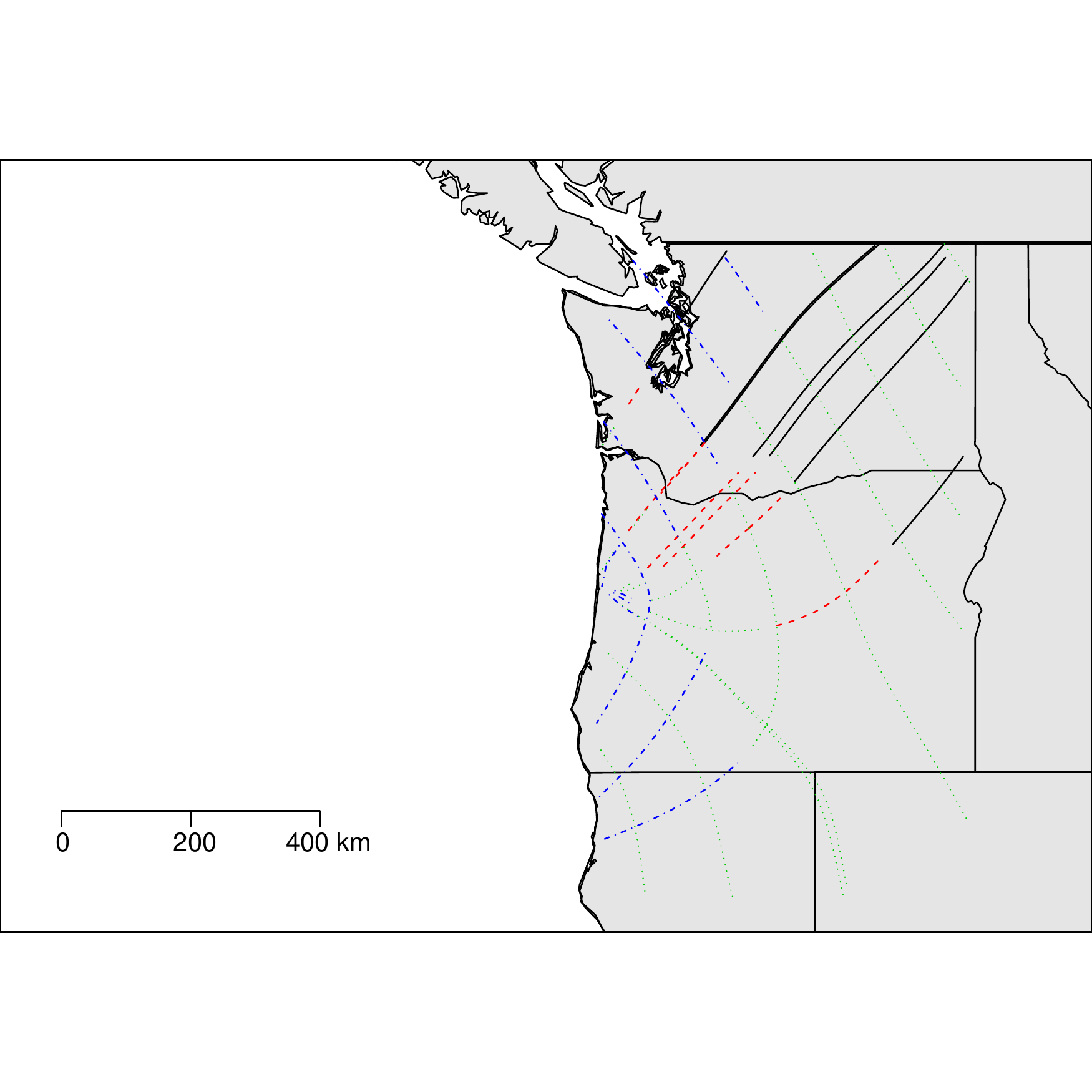}
	\end{tabular}\vspace{-.8cm}
	\caption{Biorthogonal grid for the thin-plate spline characterizing the deformation of the $\mathcal{G}$-space, using NCDC data set. Solid line indicates contraction and dashed lines indicate expansion.}
	\label{fig:Biogrid}
\end{figure}

An important diagnostic in applying the SG method, supplied with the \texttt{EnviroStat v0.4-0 R} package \citep{Le2014a}, is the biorthogonal grid seen in Figure \ref{fig:Biogrid}. It represents the degree of contracting and expanding of the $\mathcal{G}$-space needed to attain an approximately stationary domain in $\mathcal{D}$-space through deformation. The solid lines indicate contraction and dashed lines, expansion. The expansions can be explained by the abrupt changes in the residual temperatures for nearby regions, due to diverse terrain. A contraction is seen in Eastern Washington, a basin located between the Cascade and Rocky Mountains. 

\begin{figure}[h!]
	\centering
	\begin{tabular}{c}
		\includegraphics[scale=0.5]{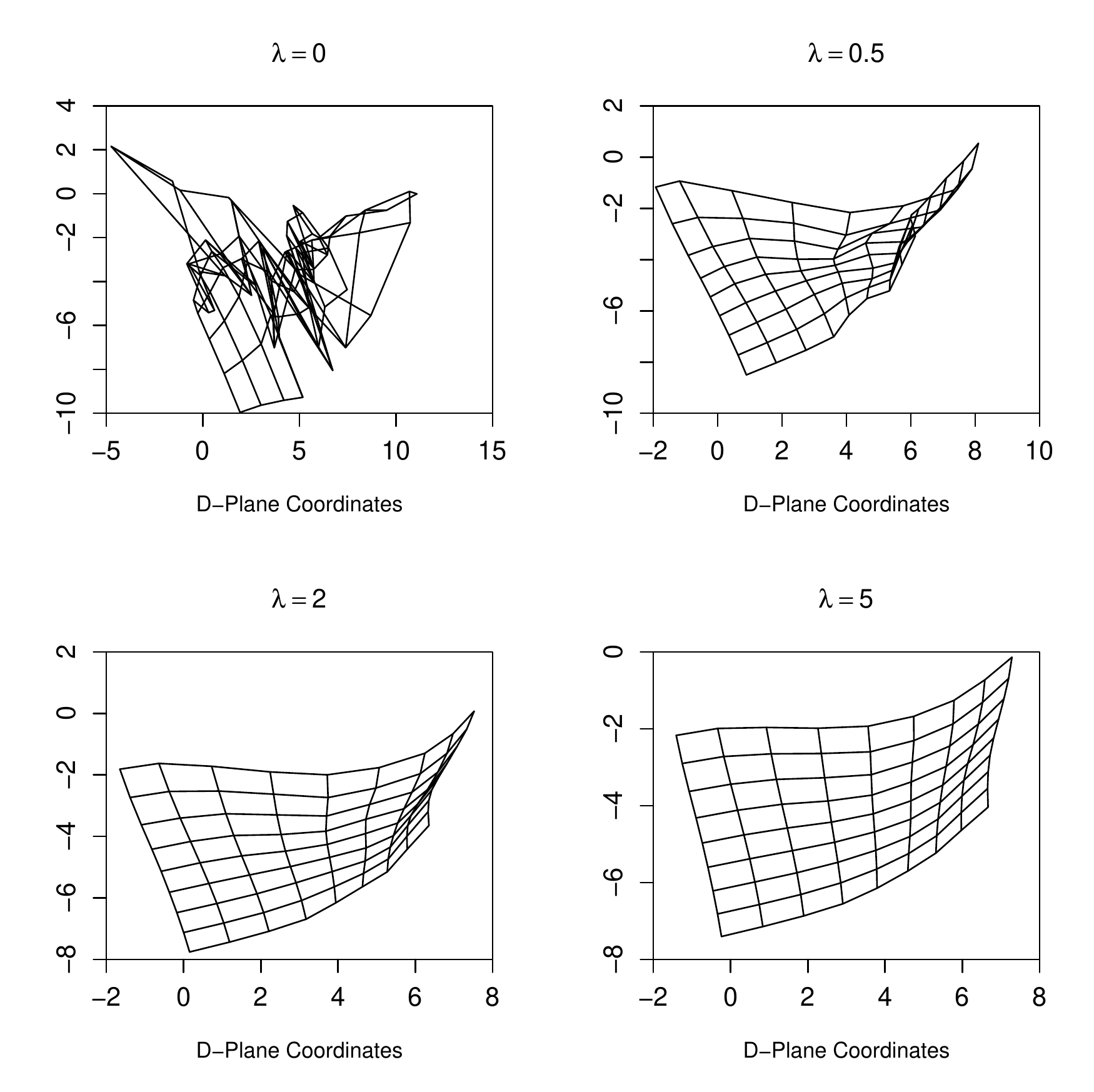}
	\end{tabular}
	\caption{Deformation assuming different spline smoothing $\lambda$ values. Note that when $\lambda = 0$, no smoothing is applied.}
	\label{fig:Dplanes}
\end{figure}

\begin{figure}[h!]
	\centering
	\begin{tabular}{cc}
		\includegraphics[scale=0.3]{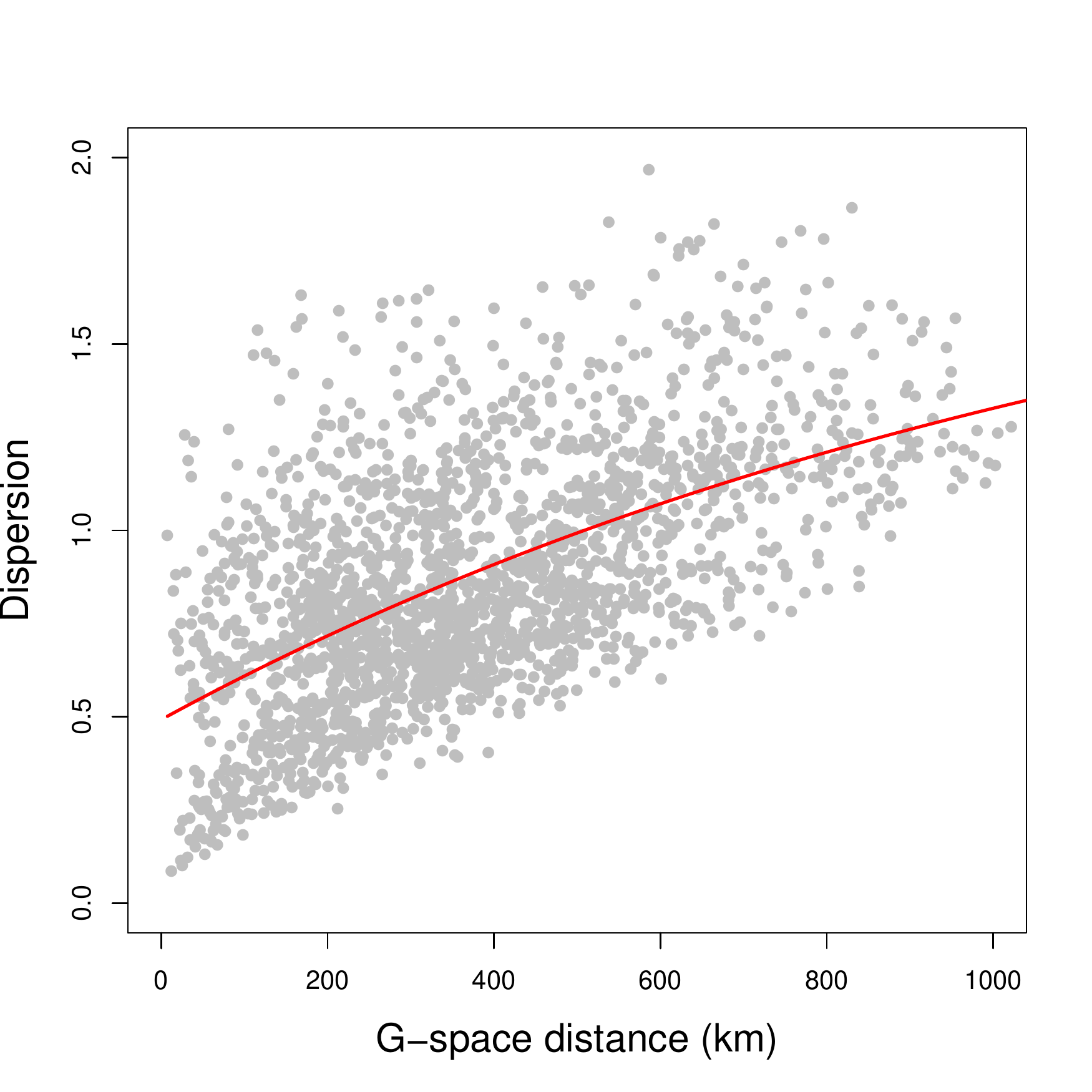}
		\includegraphics[scale=0.3]{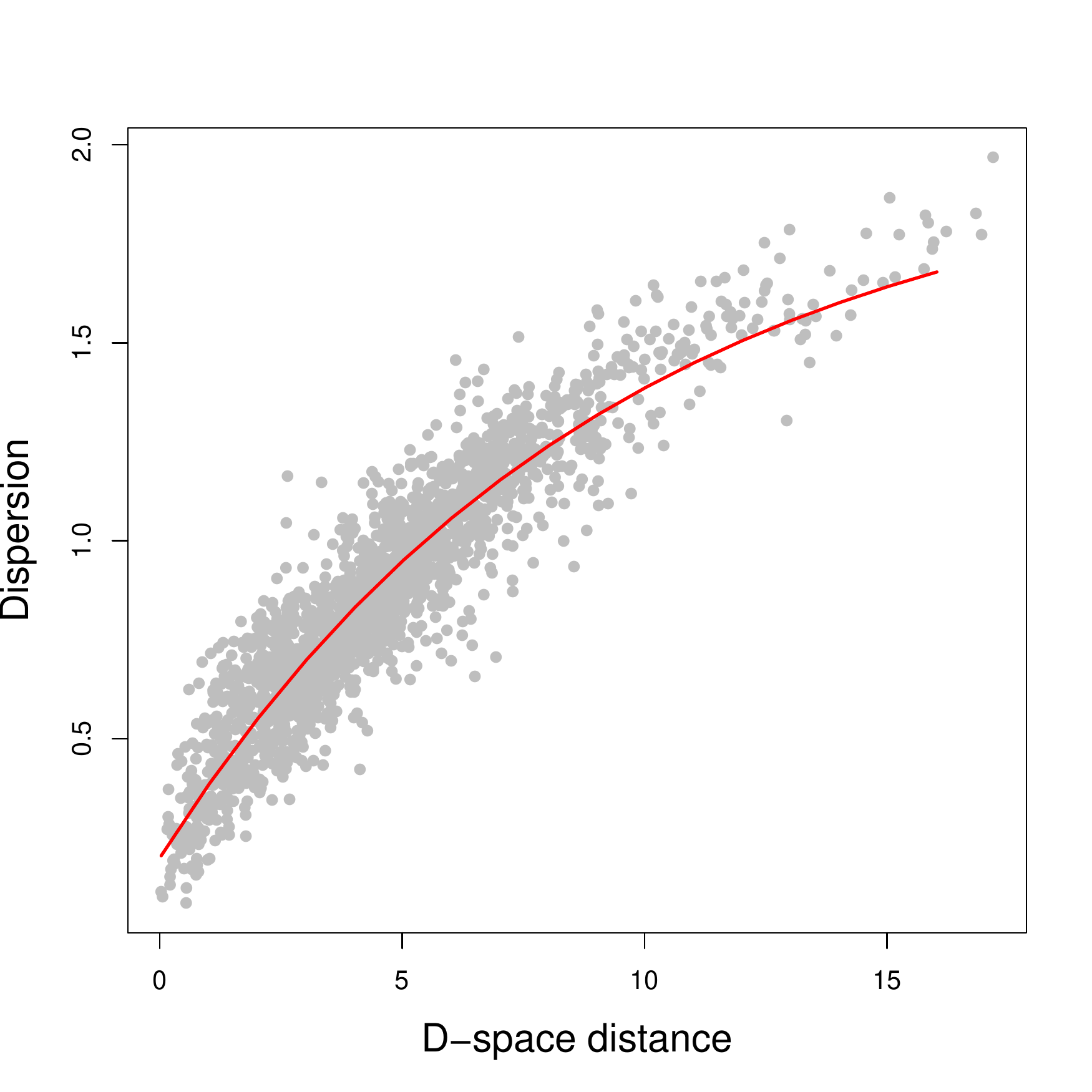}
	\end{tabular}
	\caption{Estimated dispersions after SG approach in $\mathcal{G}$-space and in $\mathcal{D}$-space. The solid line represents a fitted exponential variogram.}
	\label{fig:Dispersions}
\end{figure}

Figure \ref{fig:Dplanes} illustrates the effect of different spline smoothing $\lambda$ values in the deformed space. Without any smoothing $(\lambda = 0)$, the $\mathcal{D}$-space is folded over on itself, implying that widely separated sites tend to be more correlated than sites located between them. To make the results more interpretable, we have chosen $\lambda  = 5$, a value that keeps more of the gains from deformation seen in Figure \ref{fig:Dispersions}, without folding the $\mathcal{G}$-space.

Figure \ref{fig:Dispersions} contains estimated dispersions after SG approach \cite{Sampson1992} approach in $\mathcal{G}$-space and $\mathcal{D}$-space. A more stationary fit is seen in the distorted space, since less variability is seen around the (stationary) variogram line. 

We repeated the analysis using the PRISM data described in Section \ref{PRISM}. It corresponds to average values for temperature computed over a 30-year range (1981-2010), and we use these data to represent the expected climate in the Pacific Northwest. Instead of estimating trend coefficients based on Equation \ref{MeanFct}, we instead analyze anomalies (i.e. differences between our observed values and those via PRISM). The goal is to validate our estimated trend by comparing improvements in prediction between these different analyses. 

 \newpage

In the following section, we assess and compare the spatial predictions made by our fitted spatio--temporal model with our two alternative approaches for modelling the spatio--temporal trend. We will argue that PRISM captures well the large-scale trend, but may not capture the effects of terrain at smaller scales. 

\subsection{Spatial Prediction}\label{pred}

\begin{figure}[h!]
	\centering
	\begin{tabular}{c}
		\includegraphics[scale=0.5]{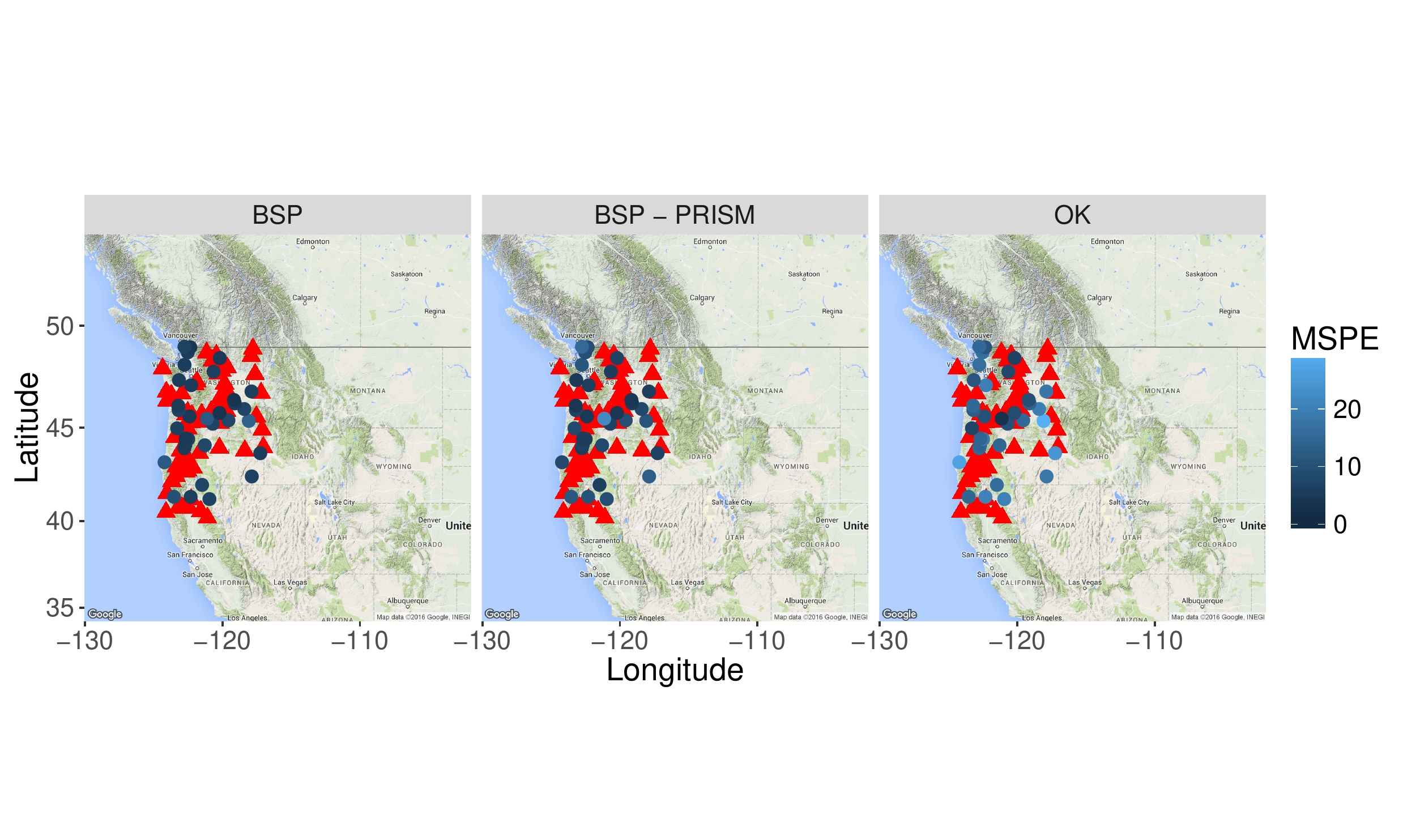}
		\end{tabular} \vspace{-1cm}
	\caption{Map of mean squared prediction errors (MSPE), averaged over time. The red triangles represent the stations used for training purposes.}
	\label{fig:MSEMap}
\end{figure}

In this section we present our assessments of the prediction accuracy of our fitted hierarchical spatio--temporal model. For validation purposes, we compare the predicted values with the real values observed for the $33$ left-out stations. Figure \ref{fig:MSEMap} contains a map of the mean squared prediction errors, averaged over time. 

\begin{figure}[h!]
	\centering
	\begin{tabular}{cc}
	\includegraphics[scale=0.5]{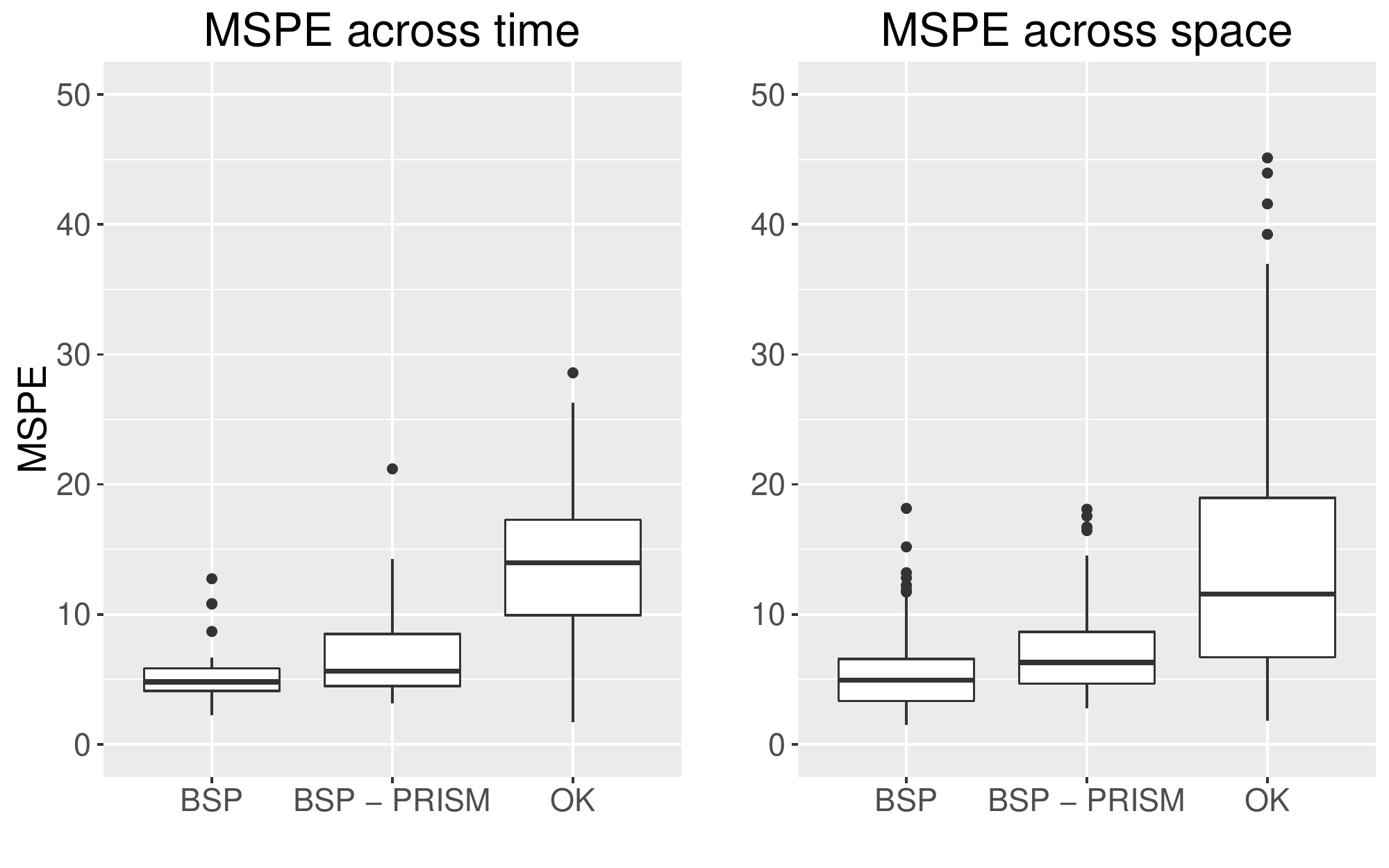}
	\end{tabular}
	\caption{Mean squared prediction errors across ungauged stations and across time for the different methods considered: Bayesian spatial prediction (BSP), Bayesian spatial prediction with PRISM (BSP - PRISM), and ordinary kriging (OK).}
	\label{fig:MSEComparison}
\end{figure}

The prediction accuracy of the hierarchical spatio-temporal Bayesian model is also compared to ordinary kriging. For ordinary kriging, we used the \texttt{geoR v.1.7-4.1 R} package \citep{RibeiroJr.2001}. The parameter estimates of the Exponential covariance function were obtained via maximum likelihood for the different time points. Figure \ref{fig:MSEComparison} displays the mean squared prediction error for the ungauged stations and for different time points, respectively. Notice that coverage for our space-time interaction spatial mean is similar when analyzing PRISM anomalies, which serves as a way to characterize the strength of our general temperature mapping theory. In addition, Figure \ref{fig:MSEComparison} shows that the mean squared prediction errors across ungauged stations and across time are, on average, smaller for the BSP method introduced considering the spatio-temporal interactions in the mean function as in Equation \ref{MeanFct}. The reason for this may be due to the fact that PRISM may not be capturing the effects of terrain at smaller scales. 

\newpage
Another disadvantage is that the PRISM data are currently not available at locations outside of the United States. Thus, we advocate that for regions with complex terrain, a thorough exploratory analysis is crucial to better understand the local changes in trend, and possible need to account for spatio-temporal interactions. 

\begin{table} \centering
\caption{Coverage probabilities and summaries for the mean squared prediction errors (MSPE) for the different methods considered: Bayesian spatial prediction (BSP), Bayesian spatial prediction with PRISM (BSP with PRISM), and ordinary kriging. The overall MSE refers to the mean squared prediction errors averaged over space and time.}
\label{tab:Sum}
\begin{tabular}{l|c|c|c}
\hline
& BSP & BSP with PRISM & Ordinary kriging \\
\hline
Coverage & 0.918 & 0.921 & 0.529 \\
Overall MSPE & 5.396 & 7.000 & 14.032 \\
Overall MSPE std. error & 2.362 & 3.733 & 5.823 \\
\hline
\end{tabular}
\end{table}

\newpage
\section{Concluding remarks}

This paper has focused on modelling temperature fields in the Pacific Northwest, where rapid changes in temperature and localized weather are common due to the complex terrain. The modelling in this region is hence rather difficult and demands flexible spatio-temporal models that are able to handle nonstationarity and those rapid changes in trend.

We introduce a flexible stochastic spatio-temporal model for daily temperatures in the Pacific Northwest that handles nonstationarity. We also stress the need for spatio-temporal interactions to understand the temperature trends. We believe that global climate models may fail to represent interesting smaller-scale trends, especially in regions with a complex terrain like the Pacific Northwest. 

We introduced two comparable strategies for spatial prediction in regions with a complex terrain. The first is an extension of the Bayesian spatial prediction \cite{Le1992} where we extended the method to take into account spatio-temporal interaction features in the mean to capture the localized changes in trend. The second is based on tackling the anomalies of the expected climate in the Pacific Northwest, based on the average values of temperature computed over a 30-year range (1981-2010), provided by PRISM Climate Group. However, we stress that the PRISM data is currently not available at locations outside of the United States. For this reason, we advocate that for regions with complex terrain, a thorough exploratory analysis is crucial to better understand the local changes in trend, and possible need to account for spatio-temporal interactions. Our work conclusively shows how appropriately modelling the spatio-temporal mean field can resolve these complex patterns for nonstationarity and improve spatial prediction. Our analysis also discovered abrupt changes in the observed temperatures for nearby regions due to diverse terrain in a great part of the western region, and less variable weather conditions in Eastern Washington, a basin located between the Cascade and Rocky Mountains. 

Ultimately the paper produces a validated method for the imputation of missing temperature measurements in the Pacific Northwest. Thus, we can use our knowledge of the behaviour of temperature fields reflected in this paper and the method to perform multiple imputation in order to deal with the highly irregular temporal spacing of the observations in the UW Probcast data set. Our ultimate goal is to combine the temperature measurements with the 48-hour forecast outputs from an ensemble of deterministic models in a statistical framework, based on the different runs of the Pennsylvania State University -- National Center for Atmospheric Research fifth generation Mesoscale Model (MM5).

\section*{Acknowledgments}
The authors would like to thank Dr. Doug Nychka for the insightful discussions and in particular for suggesting the use of the PRISM data in Section \ref{PRISM}.

% References ---------->
\bibliographystyle{imsart-nameyear}
\bibliography{references}{}

% Appendix ---------->

%%% -------------------------- >

\end{document}